\documentclass[aps,prd,nofootinbib,tightenlines,12pt]{revtex4-1}
\pdfoutput=1

\usepackage{amsmath,amssymb,amsfonts,graphicx,color}
\def\fnl{f_{\rm{NL}}}
\def\mpl{m_{\rm{pl}}}
\def\d{{\rm d}}

\begin{document}

\title{Generically large nongaussianity in small multifield inflation}

\author{Joseph Bramante}
\affiliation{Department of Physics \\ University of Notre Dame \\ Notre Dame, IN, 46556, USA}

\begin{abstract}
If forthcoming measurements of cosmic photon polarization restrict the primordial tensor-to-scalar ratio to $r < 0.01$, small field inflation will be a principal candidate for the origin of the universe. Here we show that small multifield inflation, without the hybrid mechanism, typically results in large squeezed nongaussianity. Small multifield potentials contain multiple flat field directions, often identified with the gauge invariant field directions in supersymmetric potentials. We find that unless these field directions have equal slopes, large nongaussianity arises. After identifying relevant differences between large and small two-field potentials, we demonstrate that the latter naturally fulfill the Byrnes-Choi-Hall large nongaussianity conditions. Computations of the primordial power spectrum, spectral index, and squeezed bispectrum, reveal that small two-field models which otherwise match observed primordial perturbations, produce excludably large nongaussianity if the inflatons' field directions have unequal slopes.
\end{abstract}

\maketitle

\section{Introduction}

Learning what forms of cosmological inflation are favored by observation deepens our understanding of the universe's origin and structure. This paper shows that if inflation is sourced by multiple scalar fields traversing sub-Planckian field ranges, the Planck Collaboration's bound on squeezed (aka local) nongaussianity, $\fnl^{\rm{sq}} = 2.5 \pm 5.7$ \cite{Ade:2015ava}, often implies a certain structure for small multifield potentials; without this structure they generically produce large squeezed nongaussianity.\footnote{In this work we only consider multiple small field models which terminate inflation with the dynamics of the inflating fields themselves. These considerations will apply to some (though not all) small multifield \emph{hybrid}  models \cite{Linde:1993cn}. Additionally, note that ``multifield" models as defined in this paper require that all fields source inflation over the eight observed CMB efolds, and specifically that curvaton contributions to (or damping of) primordial perturbations are not considered.} This has implications for the couplings of lifted flat directions in broken supersymmetric potentials, proposed as natural candidates for small field inflation \cite{Dine:2011ws}.

The nearly unbroken isotropy of the cosmic microwave background (CMB) provides an amusing counterexample to the well-worn notion that physicists enjoy spherical symmetry. Simply assuming that the universe began as a uniformly dense patch of radiation does not adequately explain this uniformity, which amounts to a fine-tuning of the universe's initial conditions \cite{Carroll:2014uoa}. Inflation makes this uniformity plausible, because an epoch of exponential spatial expansion (de Sitter space) fueled by a near-constant vacuum energy and preceding the radiation and matter dominated expansions of our universe, provides a common origin for otherwise causally separated lightcones on our cosmic horizon, which at present time contains primordial photons of almost equal temperature, released when the formation of hydrogen made the universe more transparent.

More concisely, the symmetry of a de Sitter space could be the source of spherical symmetry on the cosmic microwave background. But any such antecedent de Sitter space cannot be totally time symmetric: inflation must end so its effects can be observed. A simple mechanism that achieves this is a scalar field slowly-rolling down its potential, resulting in a slowly-decreasing vacuum energy that sources a quasi-de Sitter space. The necessary slowness of this roll can be quantified with the parameters
\begin{align}
\frac{\mpl^2}{2} \left(\frac{V_\phi}{V}\right)^2 &\equiv \epsilon_\phi < 1, ~~~~~~\mpl^2 \left(\frac{V_{\phi \phi}}{V}\right) \equiv \eta_\phi < 1, 
\label{eq:slowrollparams}
\end{align}
where $V(\phi)$ is a scalar potential and $V_\phi \equiv \frac{\d V }{ \d \phi}$ is the slope of the potential. The expansion fomented by this potential is measured with the scale factor of the universe, $a \propto e^{Ht}$, where $H \equiv \frac{\dot{a}}{a}$ is the Hubble rate, which can be determined from the equations of motion and continuity for a scalar field sourcing de Sitter,
\begin{align}
\ddot{\phi} + 3H \dot{\phi} + V_\phi  &= 0,~~~~~~~~~~~
\frac{1}{2} \dot{\phi}^2 +  V - 3 \mpl^2 H^2  =0, \label{eq:eomcont}
\end{align}
where in the slow-roll limit the first term in each of these equations can be omitted.

In order for our universe to be causally connected at the time of big bang nucleosynthesis, the universe needs to have expanded to roughly $e^{60}$ times its initial size during inflation, where the exponent is called the number of efolds, $N$. The expansion rate gets larger with larger vacuum energy, $V(\phi)$, and inflation lasts longer with a smaller slope, $V_\phi$. Thus it is sensible (and with Eq.~\eqref{eq:eomcont} verifiable) that in the slow-roll limit, $N$ takes the form
\begin{align}
N = \int_{t^*}^{t^e} H \d t =  \frac{1}{\mpl^2} \int_{\phi^*}^{\phi^e} \frac{V}{V_\phi} \d \phi,\label{eq:nefolds}
\end{align}
where $\phi^*$ and $\phi^e$ denote field values during and at the end of inflation. Inflation is therefore driven by a potential that is either Planck-flat or Planck-fat, more typically called small or large field inflation, meaning respectively that the slope $V_\phi$ is small compared to $\frac{V}{\mpl}$, or $\frac{V}{V_\phi} \sim \phi$ is large compared to $\mpl$. 

The latter scenario, Planck-fat inflation, occurs when in some piece of spacetime $\phi$ traverses a super-Planckian field range \cite{Linde:1983gd}. In this case, very simple potentials are sufficient to explain all present cosmological observations, for example the potential $V = m_\phi^2 \phi^2$ with initial field value $\phi^* \sim 15 ~\mpl$ and mass $m_\phi \sim 10^{-5} ~\mpl$ adequately conforms to current observations of the primordial power spectrum and spectral index \cite{Ade:2013uln,Planck:2015xua}. 

Physically, the primordial power spectrum is the result of quantum variations of the inflaton inducing variations in $H$ and equivalently in the expansion of space. If these variations decrease during inflation (we have measured a decrease) we would expect this to be evident as a decrease in the two-point correlation between CMB photon fluctuations at smaller angular separation. These photon fluctuation correlations are calculated by varying the inflaton's Einstein-Hilbert action, $S = \int d^4 x \sqrt{-g} [\frac{1}{2}R+\frac{1}{2}g^{\mu \nu} \partial_\mu \phi \partial_\nu \phi -V(\phi)]$, with respect to scalar degrees of freedom, yielding the scalar curvature perturbation and corresponding dimensionless power spectrum and spectral index,
\begin{align}
A_s = \frac{1}{24 \pi^2 \mpl^4} \frac{V}{\epsilon_\phi},~~~~~~~n_s-1 = 2 \eta_\phi -6 \epsilon_\phi,~~~~~~~A_t = \frac{2}{3 \pi \mpl^4} V. \label{eq:inflobs}
\end{align}
Here, the last equation is the dimensionless tensor power spectrum, obtained by varying the tensor degrees of freedom of the Einstein-Hilbert action. Note that $A_t$ depends only on the magnitude of $V$, because a scalar potential does not have tensor degrees of freedom. 

Despite the apparent simplicity of large field models, we should question how the small self-couplings ($m_\phi$) of a super-Planckian inflaton remain stable against radiative corrections. Without knowing the relevant UV dynamics, we should expect that terms like $\propto \frac{\phi^5}{\mpl}$, induced by integrating out heavy states that arise at the Planck scale, would disrupt the smallness of the inflaton self-coupling (e.g. $m_\phi \sim 10^{-5} ~\mpl$), when $\phi$ traverses super-Planckian field values. For large field inflation, the (clearly broken) shift-symmetries of scalar potentials may prevent these super-Planckian corrections \cite{Freese:1990rb,Silverstein:2008sg} (see also \cite{Berera:1995ie,Berera:1999ws}). Regardless of the exact mechanism, if inflation is Planck-fat, then the inflaton requires an underlying shift-symmetry, some other symmetry, or fine-tuning to enforce the smallness of its self-couplings. 

Concerns about the stability of large field models have been recently reified, because single field inflation may be necessarily Planck-fat. Note from Eq.~\eqref{eq:inflobs} that a measurement of both the scalar and tensor perturbations set the energy of the potential, $|V|$, during inflation. Using this and inserting Eqs.~\eqref{eq:inflobs} into \eqref{eq:nefolds}, it can be shown that if primordial tensor modes are measured such that $r \equiv \frac{A_t}{A_s} \gtrsim 0.01$, then single field inflation for our universe was Planck-fat. Further discussion of the Lyth Bound and its implications for $r$ can be found in Refs.~\cite{Lyth:1996im,Efstathiou:2005tq,Easther:2006qu,Baumann:2011ws,Antusch:2014cpa,Garcia-Bellido:2014eva,Gao:2014pca,Bramante:2014rva,Garcia-Bellido:2014wfa}. Most importantly, any measurement of $r$, given the sensitivity of upcoming experiments, would imply a super-Planckian field range for single field slow-roll inflation.

But if ongoing efforts to probe primordial tensor modes through the B-mode polarization of the CMB \cite{Ade:2014xna} result in a bound on tensor modes, $r \lesssim 0.01$, rather than a measurement, then symmetries that stabilize Planck-fat inflaton self-couplings would be unnecessary. Moreover, a tight enough restriction on $r$ would require small field inflation \cite{Garcia-Bellido:2014wfa}. While small field inflation would not have the self-coupling instabilities of large field inflation, there is a separate symmetry requirement for small field inflation, namely that the potential be Planck-flat. This Planck-flatness can be realized as the nearly flat directions in necessarily broken and uplifted scalar supersymmetric potentials \cite{Dine:2011ws,Dine:1995uk,Gherghetta:1995dv,Allahverdi:2006iq,Allahverdi:2006we,Enqvist:2007tf,Allahverdi:2011su,Yamada:2012tj}. This is an interesting possibility, because it indicates that the same symmetry which would resolve the weak scale hierarchy problem and allow for the unification of Standard Model gauge couplings at $\sim 10^{16} ~\rm{GeV}$, may also be responsible for cosmic inflation.

However, the perturbations produced during Planck-flat inflation have some unusual properties as compared to large field inflation. During large field inflation, additional scalar degrees of freedom can result in additional primordial scalar perturbations -- scalar fields oscillating around their minima induce isocurvature modes that can transmute to scalar curvature perturbations during reheating -- this is the curvaton mechanism \cite{Mollerach:1989hu,Linde:1996gt,Lyth:2001nq}. To wit, the variations in entropy induced by these extra oscillations are ``entropy perturbations," which can contribute to scalar curvature perturbations \cite{Kodama:1985bj,Gordon:2000hv}. Conversely, during Planck-flat inflation, these same field oscillations can damp out scalar perturbations, because the flatness of a flat supersymmetric potential depends on supersymmetric scalar fields remaining at their minima \cite{Enqvist:2011pt}. 

This paper examines another feature of Planck-flat inflation that differs from the large field picture: if small field inflation occurs along \emph{multiple} flat directions, then these flat directions will either have approximately the same slope, or produce large squeezed nongaussianity. The remainder of the introduction explains why nongaussianity is typically large in Planck-flat, but not Planck-fat, multifield inflation. Section \ref{sec:flatng} uses the Byrnes-Choi-Hall (BCH) conditions \cite{Byrnes:2008wi} to show that unequally sloped Planck-flat directions produce large squeezed nongaussianity in sum- and product-separable two-field potentials. The primordial perturbations and nongaussianity of small two-field models, which can be mapped to lifted supersymmetric flat directions, are analyzed in Section \ref{sec:examples}. Section \ref{sec:conclusion} concludes. Appendix \ref{app:deltan} gives a sketch of the $\delta N$ relation, including references necessary for reconstructing the full argument in the literature. Appendix \ref{app:formulae} contains formulae for scalar perturbations of sum- and product-separable potentials.

\subsection{Squeezed nongaussianity in large versus small multifield inflation}

\begin{figure}[t]
\includegraphics[scale=1]{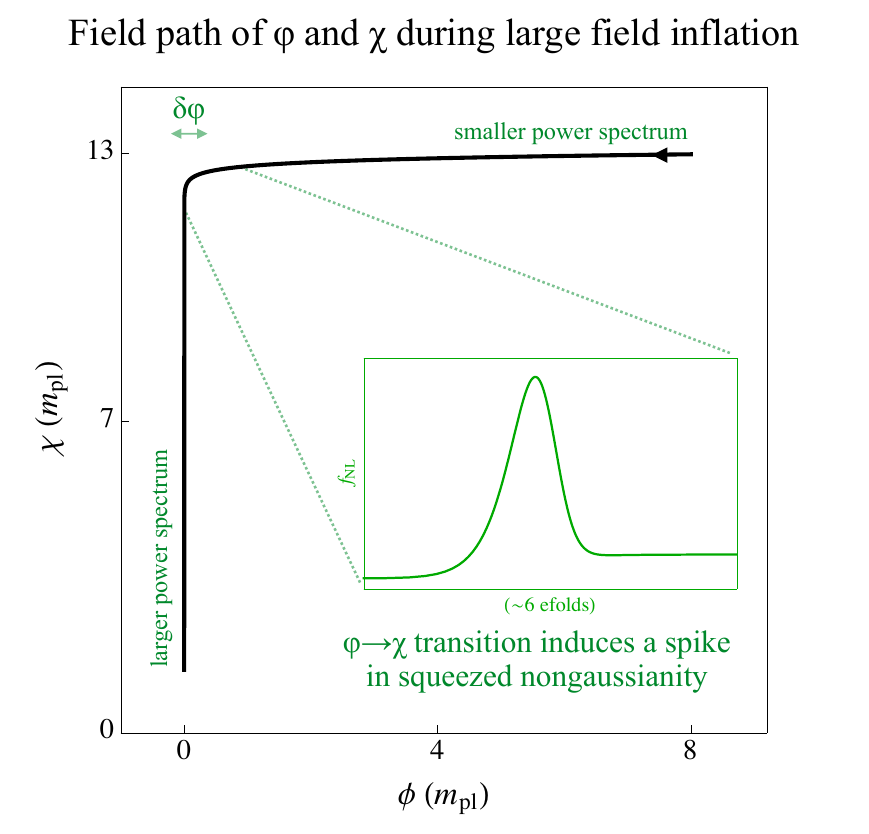}
\caption{The field space trajectory of two scalar fields with the potential $V=m_\phi^2 \phi^2 + m_\chi^2 \chi^2$ and with the masses and initial field values $(\phi^*=8~\mpl,~m_\phi=10^{-5} ~\mpl)~\rm{and}~ (\chi^*=13~\mpl,~ m_\chi=10^{-6} ~\mpl)$. At the bend in field space, there is a transition from $\phi$ to $\chi$ dominantly sourcing de Sitter vacuum energy. This transition induces a temporary spike in squeezed nongaussianity, which we have calculated following the treatment of \cite{Vernizzi:2006ve}.}
\label{fig:fieldpath}
\end{figure}

By considering the simple case of large field inflation with two scalars ($\phi, \chi$) and with the potential 
\begin{align}
V=m_\phi^2 \phi^2 + m_\chi^2 \chi^2, \label{eq:simple}
\end{align}
and using Eq.~\eqref{eq:inflobs}, we can gain some understanding of squeezed nongaussianity in both large and small multifield inflation. For concreteness, assume that $\phi$ and $\chi$ have canonical kinetic terms and the following initial field values and masses,
\begin{align} 
\phi^*=8~\mpl,&~m_\phi=10^{-5} ~\mpl, \nonumber \\ ~\rm{and}~ \chi^*=13~\mpl,&~ m_\chi=10^{-6} ~\mpl.
\end{align} 
For these parameters, because $\phi$ is much more massive than $\chi$, inflation begins with $\phi$ predominantly sourcing both de Sitter space and \emph{Gaussian} scalar curvature perturbations.\footnote{Nongaussianity is tiny, $\fnl^{\rm{sq}} = \mathcal{O}(10^{-2})$, for single field inflation \cite{Gangui:1993tt,Maldacena:2002vr}.} Inflation proceeds as though $\phi$ is the sole inflaton so long as $m_\phi^2 \phi^2 \gg m_\chi^2 \chi^2$. As inflation progresses, $\phi$ will roll down its potential and $\chi$ will take over as the effective sole inflaton when $m_\chi^2 \chi^2 \gg m_\phi^2 \phi^2$. This field space path is shown in Figure \ref{fig:fieldpath}.

During the transition between $\phi$- and $\chi$-dominated inflation, when $m_\phi^2 \phi^2 \sim m_\chi^2 \chi^2$, there is a spike of squeezed nongaussianity \cite{Vernizzi:2006ve}. This is the result of three factors: (A) the power spectrum when $\phi$ dominates is different from the power spectrum when $\chi$ dominates,\footnote{During the $\chi$-dominated portion of inflation, entropy modes from the oscillations of $\phi$ around its minimum contribute to a larger primordial power spectrum.} (B) in some spacetime regions, $\phi$'s fluctuations up and down its potential cause it to dominate for a fraction of an efold longer or shorter, (C) when $\phi$ dominates for longer or shorter, inflation lasts longer or shorter in this patch, and the power spectrum is sourced mostly by $\phi$ for longer or shorter. 

The result is that the amount of inflation sourced by $\phi$ during the $\phi \rightarrow \chi$ transition is correlated with the size of the power spectrum further along in the $\phi \rightarrow \chi$ transition. This is one way to think about nongaussianity in the squeezed limit: the power spectrum at a small scale correlates with the amount of expansion that preceded at a larger scale. On the CMB this manifests as a correlation between the size of the temperature power spectrum measured at a small angle and a temperature over- or under-density at a larger angle away from that power spectrum. From the perspective of inflaton dynamics, this simply means that regions which inflated more or less during the $\phi \rightarrow \chi$ transition end up with smaller or larger power spectra on smaller scales.\footnote{Note that large scale overdensities can correlate with either smaller or larger power on small scales, the choice of which determines whether $\fnl^{\rm{sq}}$ is positive or negative, the convention for which varies in the literature. For useful visualizations of nongaussianity, see \cite{Lewis:2011au}.}

We can now understand why large squeezed nongaussianity is uncommon in large field inflation: usually a single super-Planckian scalar field is sourcing $H$ and perturbations thereof. With one field dominating, inflation is effectively single-field and produces mostly Gaussian perturbations. Nongaussianity arises only in brief bursts during field-dominance transitions. Nevertheless, it is possible to carefully arrange the inflatons' couplings and initial field values, so that the perturbation dominance of one field gradually transitions to another over the entire course of large field inflation \cite{Byrnes:2008wi,Peterson:2010mv,Tanaka:2010km,Gong:2011cd,Elliston:2011dr,Elliston:2011et,Choi:2011me,Mazumdar:2012jj}.
Thus, while finely-tuned scenarios can source a detectable amount of squeezed nongaussianity, we generically expect tiny squeezed nongaussianity in large multifield inflation. 

On the other hand, to understand why squeezed nongaussianity can be generically large in small multifield inflation, first note that in \emph{large} multifield inflation, the energy sourced by each field depends on its super-Planckian field value, and so for most of inflation, unless the model and initial conditions are carefully tuned, $|V(\phi)| \neq |V(\chi)|$. When $|V(\phi)| \sim |V(\chi)|$, there is an associated spike in squeezed nongaussianity. Contrast this with the case of \emph{small} multifield inflation, where the potential energy remains nearly constant (i.e. $V \sim V_0$) by necessity. 

This constant potential energy arises, for example, as a result of supersymmetry-breaking terms that lift charge-neutral field directions ($\phi,\chi$) in a supersymmetric scalar potential. We can parameterize the resulting potential as,
\begin{align}
V = V(\phi,\chi) + V_0,
\label{eq:multiflat1}
\end{align}
where $V_0$ is a constant term and $V_0 \gg V(\phi,\chi)$ during inflation. We are justified in assuming that small field potentials can be put into a form with a constant vacuum energy term much larger than field-dependent terms, because in slow-roll, small field models, $V_\phi$ must be tiny compared to $\frac{V}{\mpl}$ (see Eq. \ref{eq:nefolds}). In order to source 60 efolds, the inflating fields are confined to a Planck-flat part of their potential with a nearly constant height $V_0$. 

But because $V_0$ is constant and much larger than $V(\phi,\chi)$ during inflation, the slope of each field direction in Eq. \eqref{eq:multiflat1}, rather than the magnitude of the field values as in large field inflation, will determine the slow-roll parameters and thus the scalar perturbations. If the slopes of the field directions are equal, there is no transition from one perturbation regime to another, because the perturbations are the same along each field direction, and there will be negligible squeezed nongaussianity. However, as we will show in the next section, if the slopes of two field directions differ, this results in large squeezed nongaussianity. The key point is that the conditions leading to a fleeting spike of nongaussianity (which are $\epsilon_\phi \gg \epsilon_\chi$ and $\epsilon_\chi / \epsilon_\phi \lesssim |\eta_{ \chi}-1|$) in the double quadratic model of Figure \ref{fig:fieldpath}, are satisfied along the entire inflationary field path of Planck-flat, multifield potentials with unequally sloped field directions.

\section{Generically large nongaussianity in small two-field inflation}
\label{sec:flatng}

This section demonstrates that because Planck-flat inflation requires:
\begin{itemize}
\item The slope of a scalar potential to become small in order to source a sufficient number of efolds (that is, $V_\phi \ll V/\mpl$),
\item The slope to become large for inflation to end (that is, $V_\phi \sim V/\mpl$),
\end{itemize}
these requirements, along with the constancy of vacuum energy during small field inflation, altogether imply typically large nongaussianity for sum and product-separable two-field inflation with unequally sloped Planck-flat field directions.

\subsection{Preliminary definitions}

The scalar curvature perturbation, $\zeta$, quantifies fluctuations in the past expansion of the universe,
\begin{align}
a(x)=a_0 \left(1+\zeta (x) \right),
\end{align}
where $a_0$ is the overall scale factor of the universe \cite{Lyth:2004gb}. Note that $\zeta (x)$ can be regarded simply as a scalar field quantifying how much $a(x)$ deviated from an isotropic evolution at point $x$. While this clearly defines $\zeta$, it does not indicate the relationship between $\zeta$ and the fluctuation of scalar fields during inflation. Maps of scalar field fluctuations onto $\zeta$ were greatly aided by a number of results over the last two decades \cite{Sasaki:1995aw,Sasaki:1998ug,Wands:2000dp,Lyth:2004gb,Lyth:2005fi}. These results are encapsulated in an equivalence which is often called the ``$\delta N$ formalism," namely, that the variation in number of efoldings with respect to scalar fields, is equivalent to the variation in $\zeta$. We review the derivation of this equivalence in Appendix \ref{app:deltan}.

In order to map $\zeta$ onto scalar field fluctuations, it is convenient to expand $\zeta$ into powers of Gaussian scalar fields in Fourier space,
\begin{align}
\zeta_k = \zeta_{G} (k) + \frac{3}{5} \fnl^{\rm{sq}} \int \frac{d^3 \vec{p}}{(2 \pi)^3} \zeta_G(\vec{p})\zeta_G(\vec{k}-\vec{p}), \label{eq:zetaexpansion}
\end{align}
where $\vec{k},\vec{p}$ are the momenta of perturbation modes and $\zeta_G(k)$ is a Gaussian scalar field. Because of the local properties of $\zeta_G$, this expansion has been called the ``local ansatz," and the coefficient in front of the quadratic term, the ``local nongaussianity," although it is the least spatially local of the nongaussianities \cite{Lewis:2011au}, and more accurately called squeezed nongaussianity. Hereafter we drop the ``$sq$" superscript from $\fnl$, and note that the rest of this article deals solely with squeezed nongaussianity.

Using the preceding definition of $\zeta_k$ we further define the power spectrum and squeezed bispectrum in Fourier space as,
\begin{align}
\left\langle \zeta_{\vec{k_1}} \zeta_{\vec{k_2}} \right\rangle &\equiv (2 \pi)^3 \delta^3\left(\vec{k_1}+\vec{k_2}\right)\frac{2 \pi^2}{k_1^3} A_s(k_1) 
\\
\left\langle \zeta_{\vec{k_1}} \zeta_{\vec{k_2}} \zeta_{\vec{k_3}} \right\rangle & \equiv (2 \pi)^3 \delta^3\left(\vec{k_1}+\vec{k_2}+\vec{k_3} \right) B(k_1,k_2,k_3),
\end{align}
from which we can identify the power spectrum, spectral index, and squeezed nongaussianity,
\begin{align}
n_s -1 & = \frac{ \partial \ln A_s(k_1) }{\partial \ln k_1}
\\
f_{\rm{NL}} & = \frac{5}{6} \frac{\left(k_1^3k_2^3k_3^3\right)B(k_1,k_2,k_3)}{\left(4 \pi^4 A_s^2\right)\left(k_1^3+k_2^3+k_3^3\right)}. \label{eq:cosmobs}
\end{align}
The parameter $f_{\rm{NL}}$ as identified above includes two pieces, one of which is always less than $1$, proportional to the perturbation momenta, and identified in the literature as $f_{\rm{NL}}^{(3)}$ \cite{Choi:2007su}. Because $f_{\rm{NL}}^{(3)}$ is less than unity for slowly rolling fields \cite{Lyth:2005qj,Choi:2007su}, we ignore it hereafter. The overriding $\left\langle \zeta \zeta \zeta \right\rangle$ contribution to squeezed nongaussianity when $\fnl > 1$ has the momentum dependence given in Eq.~\eqref{eq:cosmobs}.

\subsection{BCH large nongaussianity conditions applied to small two-field inflation}
 
As detailed in Appendix \ref{app:deltan}, the $\delta N$ equivalence shows that $\fnl$ is proportional to the second order variation of efolds with respect to inflating scalars. Here we will consider two real scalar fields, $\phi$ and $\chi$, so
\begin{align}
\fnl = \frac{5}{6} \frac{N_{\phi \chi} N_\phi N_\chi +N_{\phi \phi} N_\phi^2 + N_{\chi \chi}  N_\chi^2}{(N_\phi^2 + N_\chi^2)^2},
\label{eq:fnleqs}
\end{align} 
where subscripts indicate partial derivatives with respect to scalar fields, as in Eq. \eqref{eq:slowrollparams}. The fact that this formula is composed only of derivatives of $N(\phi,\chi)$ illustrates the utility of the $\delta N$ equivalence. Once $N$ is known as a function of relevant fields, all scalar perturbations can be calculated. For example, using Eq. \eqref{eq:nefolds}, one can derive this map ($N(\phi,\chi)$) for sum and product-separable scalar potentials, and then compute $\fnl$ exactly for given initial ``*" and final ``e" field values of $\phi$ and $\chi$. Appendix \ref{app:formulae} contains the exact sum and product-separable expressions for $\fnl$; the remainder of this section uses the BCH large nongaussianity conditions for these potentials, to demonstrate large nongaussianity as a generic feature of small multifield inflation. 

Using the BCH conditions, we can delineate when nongaussianity is large for both a Planck-flat sum-separable potential of the form
\begin{align}
V(\phi, \chi) &= V(\phi) + V(\chi) + V_0,
\end{align}
and a Planck-flat product-separable potential of the form
\begin{align}
V(\phi, \chi) &= V(\phi)V(\chi) + V_0,
\end{align}
where $V_0$ is nearly constant and $V_0 > V(\phi),V(\chi)$ during inflation. (To remove $V_0$ after inflation, we can either assume that the fields at their true minimum remove $V_0$ or that a hybrid mechanism is triggered to cancel $V_0$ once $\epsilon_\chi \sim \epsilon_\phi \sim 1$ \cite{Enqvist:2010vd,Yamada:2012tj}.)
A decomposition of \eqref{eq:fnleqs} yields the BCH large nongaussianity conditions for separable potentials \cite{Byrnes:2008wi,Byrnes:2008zy} in the limit $\epsilon_\chi < \epsilon_\phi$. These are
\begin{align}
\frac{\epsilon_\chi^*}{\epsilon_\chi^* + \epsilon_\phi^*} & \ll 1, 
\label{eq:sepcond1} \\
\left( \frac{\epsilon_\chi^*}{\epsilon_\chi^* + \epsilon_\phi^*} \right) & \lesssim \left( \frac{\epsilon_\chi^e}{\epsilon_\chi^e + \epsilon_\phi^e} \right)^{3/2} \left(\sqrt{|2\eta_{\chi \chi }^e-\eta_{\chi \chi }^*|} -1\right).
\label{eq:sepcond2}
\end{align}
Here we have given large nongaussianity conditions assuming $\chi$ rolls more slowly than $\phi$. Making the replacement $\phi \leftrightarrow \chi$ in \eqref{eq:sepcond1} and \eqref{eq:sepcond2} yields an equivalent set of conditions for large nongaussianity. There is a third large nongaussianity condition which reduces to \eqref{eq:sepcond1} if inflation transpires at a fixed vacuum energy \cite{Byrnes:2008zy}, as is the case for Planck-flat inflation. 

We now examine how Planck-flat, separable potentials with unequally sloped field directions typically produce large nongaussianity, given the evolution of Planck-flat slow-roll parameters and the BCH large nongaussianity conditions. Unequally sloped field directions imply $V_\chi \ll V_\phi$, and this satisfies \eqref{eq:sepcond1}. Condition \eqref{eq:sepcond2} is typically satisfied in multifield Planck-flat inflation, because of the dynamical evolution of slow-roll parameters in these models. During Planck-flat inflation, the $\eta_{\chi \chi}, \eta_{\phi \phi}$ slow-roll parameters are often greater than one, both at the onset and end of inflation, because $\epsilon_\chi$ and $\epsilon_\phi$ must become $\ll 1$ for inflation to begin, and then must become large again for inflation to end \cite{Bramante:2014rva}. Hence during Planck-flat inflation, $(2\eta_{\chi \chi}^e-\eta_{\chi \chi }^*)^{1/2}-1$ is typically $\gtrsim 1$. 

Also, the second parenthetical term in \eqref{eq:sepcond2} is usually larger than the first parenthetical term. The first large nongaussianity condition already requires that the first parenthetical term is less than unity, and it is necessary that $\phi$ and $\chi$ field directions have steep (and thus nearly equal) slopes at the end of inflation, meaning $\epsilon_\chi^e \sim \epsilon_\phi^e \sim 1$. Hence the second parenthetical term is typically $\mathcal{O}(1)$, and thus larger than the first parenthetical term, altogether satisfying the inequality of Eq.~\eqref{eq:sepcond2}.

It is clear then, that the size of nongaussianity in multifield Planck-flat inflation hinges on how differently sloped the field directions are, i.e. to what extent $V_\phi \neq V_\chi$, or equivalently in Planck-flat inflation, $\epsilon_\phi \neq \epsilon_\chi$. 

\section{The simplest small two-field potentials and generically large nongaussianity}
\label{sec:examples}

To demonstrate typically large nongaussianity in small multifield inflation, we analyze the simplest sum and product-separable two-field models. For examples of small field inflation occurring along Planck-flat field directions without a hybrid mechanism, also called inflection point inflation, see \cite{Allahverdi:2006iq,Allahverdi:2006we,Krause:2007jk,Baumann:2007np,Baumann:2007ah,Enqvist:2007tf,Baumann:2008kq,Burgess:2008ir,Allahverdi:2008bt,Chen:2009nk,Badziak:2009eh,Agarwal:2011wm,Downes:2011gi,Elliston:2012wm,Downes:2012xb,McAllister:2012am,Erdmenger:2012gx,Cerezo:2012ub,Yamada:2012tj,Downes:2012gu,Choudhury:2013jya,Pedro:2013pba}. 

Potentials containing terms that are linear and cubic in real scalar fields are the most minimal potentials for non-hybrid small field inflation, which depends on a non-vanishing third order field derivative ($V_{\phi \phi \phi } \neq 0$) and a small but non-vanishing first order field derivative ($V_{\phi} \ll V/\mpl$). These again are simply the requirements of any Planck-flat potential which does not end inflation with the hybrid mechanism. When $\Delta \phi \ll \mpl$, the slope of the potential, $V_\phi$ must become tiny so that the inflaton rolls slowly enough to source sufficient efolds. The related requirements that $V_{\phi \phi}$ is large at the outset of inflation, becomes small to let $V_\phi$ remain tiny, and then becomes large again to increase $V_\phi$ and end inflation requires $V_{\phi \phi \phi }$ to be non-vanishing. 

Scalar potentials composed only of constant, linear, and cubic terms are odd looking to a particle theorist, but they should be understood simply as the leading terms in a Taylor expansion around a lifted flat direction in a supersymmetric potential. The quadratic term will be absent because it is negligibly small; $V_{\phi \phi} \sim 0$ over the majority of the inflaton's trajectory, allowing $V_\phi$ to remain small and constant during inflation. Despite looking odd, this Taylor expansion around the flat piece of a scalar potential is the most amenable to matching onto CMB observables. 

For an example of a broken supersymmetric potential \cite{Allahverdi:2006iq,Enqvist:2010vd} with a flat segment that can be Taylor expanded into an effective inflaton potential composed of constant, linear, and cubic scalar terms, consider the gauge singlet combination of fields in the MSSM, $\Phi = udd$, where each of the latter is a chiral superfield. This gauge singlet combination is known as a monomial, and all monomials and higher order effective operators that break the flatness of, e.g. $udd$ in the MSSM, have been classified in \cite{Gherghetta:1995dv}. These appear in the superpotential as $W \supset \lambda \Phi^n / M^{n-3}$. If we define $\phi$ as the scalar part of the gauge invariant degree of freedom contained in $\Phi=udd$, then leading order terms that break the superpotential's $\phi \rightarrow \phi + \phi_0$ invariance are
\begin{align}
V \supset \frac{m^2}{2} |\phi|^2 + \frac{A \lambda}{n M^{n-3}} |\phi|^n + \frac{|\lambda|^2}{ M^{2n-6}} |\phi|^{2n-2},
\label{eq:susybreakingterms}
\end{align}
where $m$ and $A$ are the usual soft mass and trilinear coupling of supersymmetry-breaking soft terms. As demonstrated in \cite{Allahverdi:2006iq}, if 
\begin{align}
A \simeq (8n-8)^{1/2}m,
\end{align}
then at some point in the potential, $\phi_0$, there will be a flat segment or ``inflection point," where $V_{\phi \phi}(\phi_0) \sim 0$. Around this flat segment it is useful to define the Taylor expansion of the potential in $\phi$,
\begin{align}
V_T(\phi) = V(\phi_0) + V_{\phi}(\phi_0) (\phi-\phi_0)+ V_{\phi \phi \phi}(\phi_0) (\phi-\phi_0)^3,
\end{align}
where again the omitted quadratic term can be neglected. To simplify notation in the work that follows, we define $\lambda_1 = V_{\phi}(\phi_0)$, $\lambda_3 = V_{\phi \phi \phi}(\phi_0)$, $\gamma_1 = V_{\chi}(\chi_0)$, and $\gamma_3 = V_{\chi \chi \chi}(\chi_0)$, and re-center the Taylor expansion at $\phi_0=\chi_0=0$. With these conventions, we build small two-field potentials composed of a vacuum energy term $V_0$, along with either the sum or product of $(\lambda_3 \phi^3 + \lambda_1 \phi)$ and $(\gamma_3 \chi^3 + \gamma_1 \chi)$. We will see that, after $\gamma_3,\lambda_3, V_0$, and initial field values have been fixed by the requirement that inflation lasts 60 efoldings, nongaussianity is excludably large unless $\epsilon_\phi = \epsilon_\chi$, for parameter space which otherwise matches CMB observations.

\subsection{Large nongaussianity in uncoupled small two-field inflation}

We first consider a small multifield potential of the form $V(\phi,\chi) = V(\phi)+V(\chi)+V_0$,
\begin{align}
V(\phi , \chi) =  \lambda_1 \phi + \lambda_3 \phi^3 + \gamma_1 \chi + \gamma_3 \chi^3    +V_0 \label{eq:sumpot}.
\end{align}
As explained in the beginning of Section \ref{sec:examples}, this should be physically understood as the Taylor expansion of a two-field potential, at the point where the potential is flat in the $\chi$ and $\phi$ directions, making it suitable for small multifield inflation. The quadratic terms can be omitted, because they are necessarily tiny during small field inflation.

In multifield sum-separable potentials, the number of efoldings is additive,
\begin{align}
N =  \frac{1}{\mpl^2} \left( \int_{\phi^*}^{\phi^e} \frac{\lambda_3 \phi^3 + \lambda_1 \phi + V_0/2}{3\lambda_3 \phi^2 + \lambda_1} d \phi +\int_{\chi^*}^{\chi^e} \frac{\gamma_3 \chi^3 + \gamma_1 \chi + V_0/2}{3\gamma_3 \chi^2 + \gamma_1} d\chi \right) \equiv N_\phi + N_\chi, \label{eq:nefoldssum}
\end{align}
and so we define the number of efolds sourced by $\phi$ and $\chi$ as $N_\phi$ and $N_\chi$. For simplicity, we will assume that each field sources half the necessary efolds, so that $N_\phi=N_\chi=30$. While one might also consider the scenario where each field sources a different number of efolds, in order to be observed on the CMB or in large scale structure surveys, both fields must be slowly-rolling and sourcing perturbations over the first 10 efolds of inflation \cite{Bramante:2014rva}.

Using typical parameters for a single field model \cite{Allahverdi:2006iq,Enqvist:2010vd}, assuming supersymmetry is broken at scale $M \sim ~\rm{TeV}$, we set
\begin{align}
\lambda_3=\gamma_3= 10^{-17} ~\mpl, ~~~~~~~~ \lambda_1 \sim \gamma_1 \sim ( 3 \times 10^{-13} ~\mpl)^3, ~~~~~~~~ 
V_0 = (6 \times 10^{-7} ~\mpl )^4  \nonumber,
\end{align}
where we will vary $\lambda_1$ and $\gamma_1$ to change the slope of the potential in these field directions. 

While models could be considered where $\lambda_3 \neq \gamma_3$, this would require varying all four of $\lambda_1,\gamma_1,\lambda_3,\gamma_3$ to determine the behavior of the potential for differently sloped field directions, and does not qualitatively change the result. 

The condition that inflation ends when $\epsilon_\phi^e = \epsilon_\chi^e \simeq  1$ determines $\phi^e = \chi^e \simeq - 10^{-5} ~\mpl$. Because we have centered the potential around the origin of each field direction, each field will begin at a field value $\lesssim 10^{-5} ~\mpl$ and transit a maximum field range of $2 \times 10^{-5} ~\mpl$ during inflation. With all other parameters fixed as indicated, for each choice of $\lambda_1$ and $\gamma_1$, we solve for $\phi^*$ and $\chi^*$ so that each field sources 30 efolds during inflation. Note that with all parameters determined and $\lambda_3=\gamma_3$, the slopes of the potential along each field direction, $V_\phi$ and $V_\chi$, are uniquely determined by $\lambda_1 $ and $\gamma_1$. For example, if $\lambda_1 < \gamma_1$ the field direction along $\phi$ will be flatter than that of $\chi$, and because each is required to source 30 efolds, this mandates $\phi^* < \chi^*$. With this procedure we examine primordial scalar perturbations when one field direction is more sloped than the other. 

\begin{figure}
\center
\includegraphics[width=.99\textwidth]{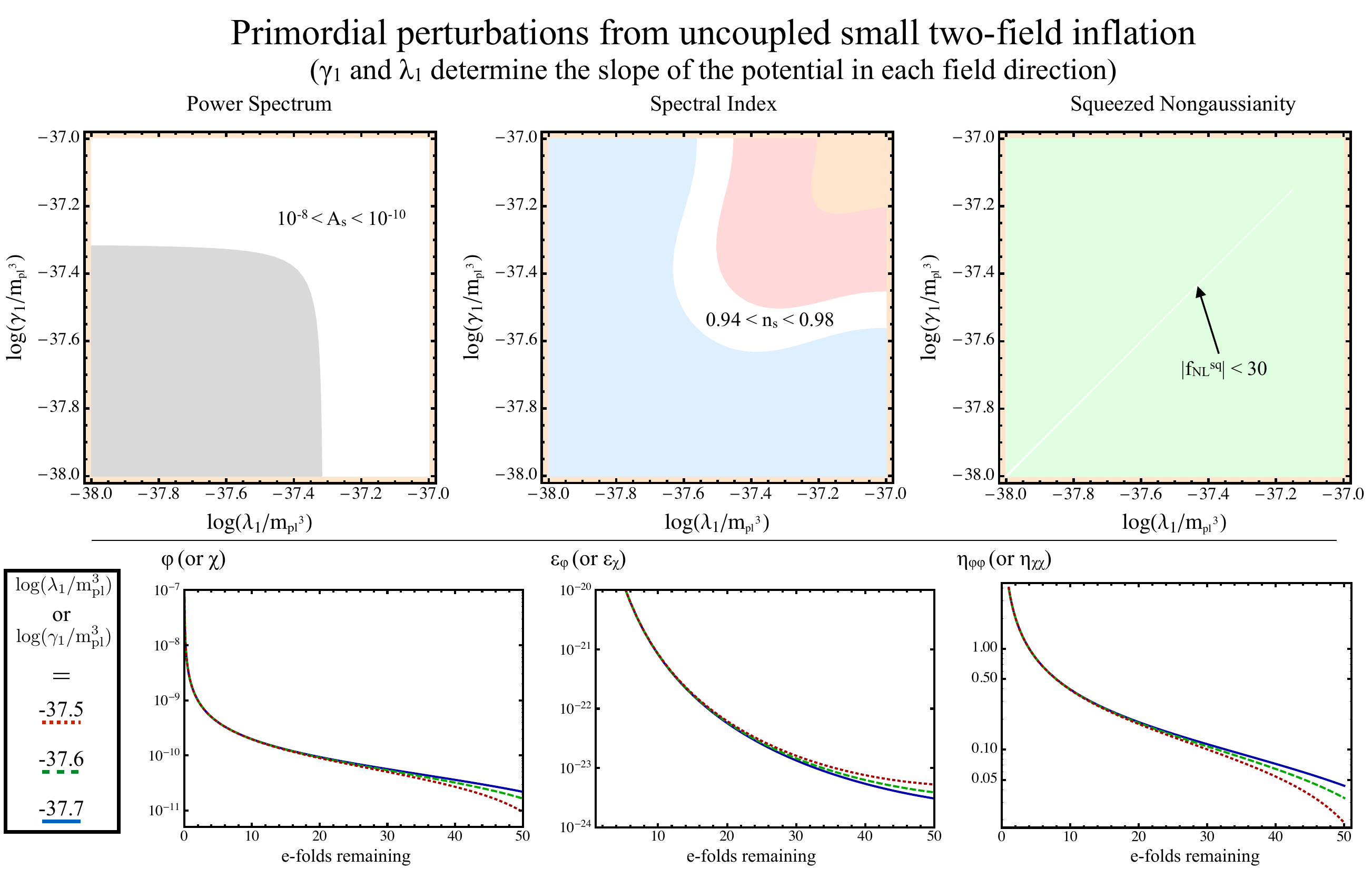}
\caption{{\bf Top:} Parameter space scans for sub-Planckian, sum-separable two-field inflation are shown in terms of $\gamma_1$ and $\lambda_1$, which determine the slope of the potential along each field direction (i.e. $V_\phi$ and $V_\chi$). All other parameters have been fixed by requiring 30 efolds of inflation along each field direction when $H \sim V_0 \sim 10^{12}~\rm{GeV}$ (see Eq. \ref{eq:sumpot} and discussion). While differently sloped field directions do not restrict the power spectrum and spectral index, Planck's bound on nongaussianity ($\fnl^{\rm{sq}} = 2.5 \pm 5.7$) requires the field directions have equal slopes. (Left panel) Grey shaded regions indicate $A_s > 10^{-10}$ or $A_s < 10^{-8}$. (Center panel) Blue shaded regions indicate $n_s > 0.98$, red indicate $n_s < 0.94$. (Right panel) Note the thin white line of parameter space, passing the requirement $|\fnl^{\rm{sq}}| < 30$. In all three panels, orange shaded regions were not scanned. {\bf Bottom:} The evolution of fields and slow roll parameters are shown for $\phi$ and $\chi$, for a number of values of $\lambda_1,\gamma_1$. Note that $\eta_{\phi \phi}$ and $\eta_{\chi \chi}$ are $\mathcal{O}(1)$ at the end of inflation.}
\label{fig:sum}
\end{figure}

We plot the resulting power spectrum, spectral index, and squeezed nongaussianity in Figure \ref{fig:sum}. Analytic expressions for these observables can be found in Appendix \ref{app:formulae}. From Figure \ref{fig:sum}, we can surmise that the relative flatness of field directions in two-field, uncoupled, Planck-flat potentials is rigidly constrained. While there is ample parameter space with $\lambda_1 \neq \gamma_1$ and consequently $V_\phi \neq V_\chi$ that is consistent with observations of the power spectrum and spectral index, in order to be consistent with the Planck constraint on squeezed nongaussianity of $\fnl = 2.7 \pm 5.7$, it is required that $\lambda_1 \sim \gamma_1$ and $V_\phi \sim V_\chi$.

\subsection{Large nongaussianity in coupled small two-field inflation}

The preceding section demonstrated large nongaussianity for unequally sloped field directions in a sum-separable potential. We will now consider the simplest product-separable small field model of the form $V(\phi,\chi) = V(\phi)V(\chi) + V_0$,
\begin{align}
V(\phi,\chi) = \frac{1}{\mpl^4} (\lambda_1 \phi + \lambda_3 \phi^3 + V_0)( \gamma_1 \chi + \gamma_3 \chi^3 + V_0). \label{eq:toyproduct}
\end{align}
While this potential contains only non-renormalizable terms, it is the simplest product-separable Planck-flat potential, and we  use it to study the scalar perturbations of two small field inflatons that are significantly coupled.

\begin{figure}
\center
\includegraphics[width=0.99\textwidth]{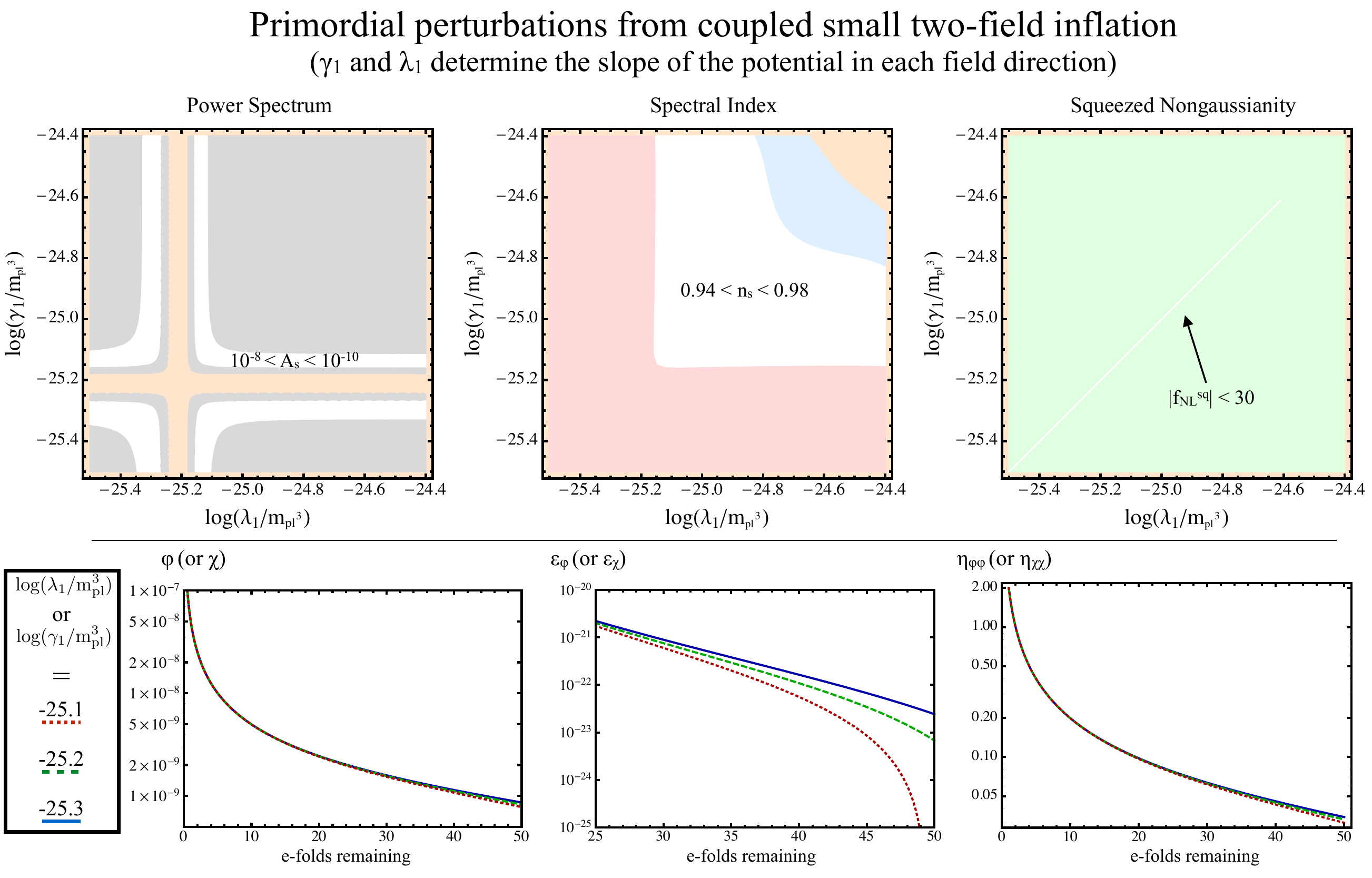} \\
\caption{{\bf Top:} Parameter space scans for sub-Planckian, product-separable two-field inflation are shown in terms of $\gamma_1$ and $\lambda_1$, which determine the slope of the potential along each field direction (i.e. $V_\phi$ and $V_\chi$). All other parameters have been fixed by requiring 30 efolds of inflation along each field direction when $H \sim 10^{11}~\rm{GeV}$ (see Eq. \ref{eq:toyproduct} and discussion). While the power spectrum and spectral index can fit CMB and LSS measurements for a variety of $\gamma_1$, $\lambda_1$ values, Planck's bound on nongaussianity ($\fnl^{\rm{sq}} = 2.5 \pm 5.7$) requires the field directions have equal slopes. (Left panel) Grey shaded regions indicate $A_s > 10^{-10}$ or $A_s < 10^{-8}$. (Center panel) Blue shaded regions indicate $n_s > 0.98$, red indicate $n_s < 0.94$. (Right panel) Note the thin white line of parameter space allowed after requiring $|\fnl^{\rm{sq}}| < 30$. In all three panels, orange shaded regions were not scanned. {\bf Bottom:} The evolution of fields and slow roll parameters are shown for $\phi$ and $\chi$, for a number of values of $\lambda_1,\gamma_1$. Note that $\eta_{\phi \phi}$ and $\eta_{\chi \chi}$ are $\mathcal{O}(1)$ at the end of inflation.}
\label{fig:prod}
\end{figure}

For product-separable potentials, the total number of efoldings is equal along each field direction,
\begin{align}
N = \frac{1}{\mpl^2} \int_{\phi^*}^{\phi^e} \frac{ \lambda_3 \phi^3 + \lambda_1 \phi + V_0}{3\lambda_3 \phi^2 + \lambda_1}d \phi =\frac{1}{\mpl^2} \int_{\chi^*}^{\chi^e} \frac{\lambda_3 \chi^3 + \lambda_1 \chi + V_0}{3\lambda_3 \chi^2 + \lambda_1} d \chi \equiv N_\phi = N_\chi ,\label{eq:nefoldsprod}
\end{align}
and so we require that $N_\phi = N_\chi = 60$. Choosing parameters that produce a power spectrum and spectral index in line with observation, we set
\begin{align}
\lambda_3=\gamma_3= 10^{-7} ~\mpl, ~~~~~~~~ \lambda_1 \sim \gamma_1 \sim ( 3 \times 10^{-9} ~\mpl)^3, ~~~~~~~~ 
V_0 = (3 \times 10^{-4} ~\mpl )^4.  \nonumber
\end{align} 
Note that with these choices, the total vacuum energy of the potential is $V(\phi,\chi) \simeq V_0^2 / \mpl^4 = (10^{-7} ~\mpl)^4$ during inflation. As before, the requirement that $\epsilon_\phi^e = \epsilon_\chi^e \simeq  1$ determines $\phi^e = \chi^e \simeq - 10^{-4} ~\mpl$. Each choice of $\gamma_1$ and $\lambda_1$, and equivalently the slope of the field direction in the $\phi$ and $\chi$ directions, fixes the initial field values $\phi^*$ and $\chi^*$ such that 60 efolds of inflation are produced. Varying $\lambda_1$ and $\gamma_1$, we compute the power spectrum, spectral index, and squeezed nongaussianity, using the formulae in Appendix \ref{app:formulae}.

In Figure \ref{fig:prod}, we see that coupling between the $\phi$ and $\chi$ field directions in small two-field inflation leads us to the same conclusions as in the uncoupled case. While potentials with unequally sloped field directions can match observations of the primordial power spectrum and spectral index, squeezed nongaussianity is excludably large unless $\gamma_1 \sim \lambda_1$ and equivalently $V_\phi \sim V_\chi$ along the field path of the inflatons.

\section{Conclusions}
\label{sec:conclusion}
This article has identified a feature of small multifield inflation, that nongaussianity is typically large unless small field directions have equal slopes. We have examined how nongaussianity differs in large versus small field models, applied the BCH large nongaussianity conditions to two-field models with sub-Planckian field excursions, and computed the power spectrum, spectral index, and nongaussianity for the simplest examples of coupled and uncoupled small two-field inflation. Results from each of these indicate that unless the slopes of field directions in a small multifield potential are equal, large nongaussianity results. This can be contrasted with large multifield inflation, in which different initial field values and slopes do not typically induce large nongaussianity.

This result has implications for patterns of supersymmetry breaking that permit small field inflation along the gauge invariant field directions of a once supersymmetric potential. Either inflation occurs over a single field direction, which requires that only one set of supersymmetric scalar fields become excited to sub-Planckian energies at the onset of inflation (e.g. $\Phi = udd$ and not $\Phi = css$ transits a sub-Planckian field range), or supersymmetry is broken such that each participating sparticle generation has equal soft masses and trilinear couplings, so that the field directions will have equal slopes in the resulting potential.

While in large field inflation, it is necessary to invoke symmetries to stabilize the inflaton against corrections to its self-couplings as it traverses a super-Planckian field range, in small field inflation, it is not enough to assume inflation is supported by the residual symmetries of a broken supersymmetric potential. Rather there is an additional requirement: either inflation must occur along just one flat direction, or the flat directions on which it occurs must be equally sloped.

\acknowledgments
I thank Christian Byrnes for useful comments on the manuscript and pointing out a factor of two error in Eq. \ref{eq:nefoldssum}. I also thank the anonymous referee for suggestions to improve the manuscript, along with Latham Boyle, Chris Brust, Jessica Cook, Landon Lehman, Adam Martin, Subodh Patil, Raquel Ribeiro, David Seery, and Kendrick Smith for useful discussions, and Sean Downes for early collaboration. I am grateful to the CERN theory division for hospitality while portions of this work were completed. This research was supported in part by Perimeter Institute for Theoretical Physics. Research at Perimeter Institute is supported by the Government of Canada through Industry Canada and by the Province of Ontario through the Ministry of Economic Development \& Innovation. 

\appendix
\section{Deriving the $\delta N$ equivalence}
\label{app:deltan}

This appendix sketches out the equivalence between comoving scalar curvature perturbations during inflation and perturbations to the number of efoldings, a relationship often called the $\delta N$ formalism. We hew closely to the treatment of \cite{Sasaki:1995aw}. Consider the most general form of linear scalar perturbations to a de Sitter metric,
\begin{align}ds^2 = -(1+2A)dt^2 + 2 \delta_iB dx^idt-a^2(t)\left[(1+2 \zeta)\delta_{ij} + 2 E\delta_{i}\delta_j \right]dx^i dx^j. \label{eq:desittpert}
\end{align}
Inspecting Eq. \eqref{eq:desittpert}, it is clear that $\zeta$ is the traceless scalar curvature perturbation of a constant time hypersurface. If we define a volume expansion rate $\theta \equiv v^\mu_{;\mu}$ of constant time hypersurfaces for this perturbed de Sitter space, and define its total expansion over a path $f(\tau)$ as
\begin{align}
\mathcal{N} \equiv \int_{f(\tau)} \frac{1}{3} \theta d\tau,
\end{align}
where $\tau$ is conformal time, then it has been shown that if one also assumes that anisotropic stress perturbations are small, this volume expansion rate is given by (see Ref.~\cite{Kodama:1985bj}),
\begin{align}
\frac{1}{3} \theta \simeq H\left(1-A+\frac{1}{H} \dot{\zeta}\right).
\end{align}
The key point is that assuming small anisotropic stress removes the off-diagonal parameters present in Eq.~\eqref{eq:desittpert} from $\theta$. Given this expression for $\theta$, and using the identity $d \tau = (1+A) dt$, where $d \tau^2$ is simply the first term on the right side of Eq. \eqref{eq:desittpert}, and discarding terms of order $A^2$,
\begin{align}
\mathcal{N} = \int_{f(\tau)} H\left(1-A+\frac{1}{H} \dot{\zeta}\right) d \tau = \int_{f(\tau)} H\left(1-A+\frac{1}{H} \dot{\zeta}\right) (1+A)dt.
\end{align}
Again dropping terms proportional to the square of perturbations (that is, $A^2,A\dot{\zeta}$), this indicates a trivial relationship between the variation of efoldings and scalar curvature perturbation,
\begin{equation}
\delta N \equiv (\mathcal{N} - N) |_{f(\tau)} = \zeta|_{f(\tau)}.
\end{equation}
In summary, the perturbations to the number of efoldings generated along the inflationary field path are equivalent to the scalar curvature perturbation, in the absence of anisotropic stress perturbations, which would be nearly absent if only scalar fields source de Sitter. Thus, if inflation is sourced by scalars that roll slowly, the scalar perturbations resulting can be found by taking the variation of $N$ with respect to inflating scalar fields. Therefore,
\begin{equation}
\zeta = \delta N = \sum_a N_a \delta \phi_a^*,
\end{equation}
where $N_{a} \equiv \delta N/\delta \phi_a$ and ``*" indicates evaluation at some point along the inflationary trajectory in the scalar field potential, or equivalently, some time $t^*$ when a scalar perturbation exits the Hubble horizon during inflation.

\section{Scalar perturbation formulae for sum and product-separable potentials}
\label{app:formulae}

\subsection{Sum-separable formulae}
To simplify expressions for cosmological observables, we first define $\epsilon \equiv \epsilon_\phi+\epsilon_\chi$.
For sum-separable scalar potentials in particular it is useful to define the following set of parameters \cite{Vernizzi:2006ve},
\begin{align}
Z^e \equiv \frac{(V(\chi)^e+V_0) \epsilon^e_\phi- (V(\phi)^e +V_0) \epsilon^e_\chi}{\epsilon^e},
~~~~ u \equiv \frac{V(\phi)^*+V_0+Z^e}{V(\phi,\chi)^*},~~~~ v \equiv \frac{V(\chi)^*+V_0- Z^e}{V(\phi,\chi)^*}, \nonumber \\
\label{eq:sumsepparams}
\end{align}
where ``*" superscripts indicate terms evaluated at some point along the inflationary trajectory, ``e" superscripts indicate terms evaluated at the end of inflation, and $V(\phi)$ and $V(\chi)$ indicate the $\phi$ and $\chi$ parts of the sum-separable potential.

There are a few additional parameters that simplify the expressions for sum-separable scalar perturbations,
\begin{align}
~~~~~\eta^e_{ss} \equiv \frac{\eta_\chi^e \epsilon_\phi^e+\eta_\phi^e \epsilon_\chi^e}{\epsilon^e},~~~~~\mathcal{A}_S \equiv -\left( \frac{V(\phi,\chi)^e}{V(\phi,\chi)^*} \right)^2  \frac{\epsilon^e_\phi \epsilon^e_\chi}{\epsilon^e}
\left(1- \frac{\eta^e_{ss}}{\epsilon^e}\right).
\end{align}

In terms of these and prior variables, the power spectrum, spectral index, and $f_{NL}$ are,
\begin{align}
A_s^* &= \frac{V(\phi,\chi)^*}{24 \pi^2 \mpl^4}  \left(\frac{u^2}{\epsilon^*_\phi}   +   \frac{v^2}{\epsilon^*_\chi}     \right)
\\
n_s^*-1 &= -2 \epsilon_* - 4 \frac{u\left(1- \frac{\eta^*_\phi}{2 \epsilon^*_\phi} u\right)
                      				  +v\left(1- \frac{\eta^*_\chi}{2 \epsilon^*_\chi} v\right)   }
				  {\frac{u^2}{\epsilon^*_\phi}   +   \frac{v^2}{\epsilon^*_\chi}}
\\
\frac{6}{5}\fnl^*  &= 2 \frac{
\frac{u^2}{ \epsilon^*_\phi} \left(1- \frac{\eta^*_\phi}{2 \epsilon^*_\phi} u\right)
+\frac{v^2}{ \epsilon^*_\chi} \left(1- \frac{\eta^*_\chi}{2 \epsilon^*_\chi} v\right)
+ \left(\frac{u}{ \epsilon^*_\phi}- \frac{v}{ \epsilon^*_\chi}\right)^2 \mathcal{A}_S}
{\left(\frac{u^2}{\epsilon^*_\phi}   +   \frac{v^2}{\epsilon^*_\chi}\right)^2}
\end{align}

\subsection{Product-separable formulae}

After defining a few simplifying terms \cite{Choi:2007su},
\begin{align}
\eta^e_{ss} &\equiv \frac{\eta_\chi^e \epsilon_\phi^e+\eta_\phi^e \epsilon_\chi^e}{\epsilon^e},~~~~
\mathcal{A}_P \equiv -\frac{\epsilon_\phi^e \epsilon^e_\chi}{(\epsilon^e)^2}  
\left(\eta_{ss}^e-4 \frac{\epsilon_\phi^e \epsilon^e_\chi}{\epsilon^e}\right),
\nonumber\\
u & \equiv \frac{\epsilon^e_\phi}{\epsilon^e},~~~~~~~~~~~~~~~~~~ v \equiv  \frac{\epsilon_\chi^e}{\epsilon^e},~~~~
\end{align}
the expressions for the power spectrum, spectral index, and non-Gaussianity at lowest order in the $\delta$N formalism are,
\begin{align}
A_s^* &= \frac{V(\phi,\chi)^*}{24 \pi^2 \mpl^4}  \left(\frac{u^2}{\epsilon^*_\phi}   +   \frac{v^2}{\epsilon^*_\chi}     \right) \label{eq:powerprod}
\\
n_s^*-1 &= -2 \epsilon^* - 4 \frac{u^2\left(1- \frac{\eta^*_\phi}{2 \epsilon^*_\phi} \right)
                      				  +v^2\left(1- \frac{\eta^*_\chi}{2 \epsilon^*_\chi} \right)   }
				  {\frac{u^2}{\epsilon^*_\phi}   +   \frac{v^2}{\epsilon^*_\chi}}\label{eq:specprod}
\\
\frac{6}{5}f_{NL}^{*}  &= 2 \frac{
\frac{u^3}{ \epsilon^*_\phi} \left(1- \frac{\eta^*_\phi}{2 \epsilon^*_\phi} \right)
+\frac{v^3}{ \epsilon^*_\chi} \left(1- \frac{\eta^*_\chi}{2 \epsilon^*_\chi} \right)
- \left(\frac{u}{ \epsilon^*_\phi}- \frac{v}{ \epsilon^*_\chi}\right)^2 \mathcal{A}_P}
{\left(\frac{u^2}{\epsilon^*_\phi}   +   \frac{v^2}{\epsilon^*_\chi}\right)^2},\label{eq:fnlprod}
\end{align}


\bibliographystyle{JHEP.bst}

\bibliography{multiNotes}

\providecommand{\href}[2]{#2}\begingroup\raggedright\begin{thebibliography}{10}

\bibitem{Ade:2015ava}
{\bf Planck Collaboration} Collaboration, P.~Ade et~al., {\it {Planck 2015
  results. XVII. Constraints on primordial non-Gaussianity}},
  \href{http://arxiv.org/abs/1502.0159}{{\tt arXiv:1502.0159}}.

\bibitem{Linde:1993cn}
A.~D. Linde, {\it {Hybrid inflation}},  {\em Phys.Rev.} {\bf D49} (1994)
  748--754, [\href{http://arxiv.org/abs/astro-ph/9307002}{{\tt
  astro-ph/9307002}}].

\bibitem{Dine:2011ws}
M.~Dine and L.~Pack, {\it {Studies in Small Field Inflation}},  {\em JCAP} {\bf
  1206} (2012) 033, [\href{http://arxiv.org/abs/1109.2079}{{\tt
  arXiv:1109.2079}}].

\bibitem{Carroll:2014uoa}
S.~M. Carroll, {\it {In What Sense Is the Early Universe Fine-Tuned?}},
  \href{http://arxiv.org/abs/1406.3057}{{\tt arXiv:1406.3057}}.

\bibitem{Linde:1983gd}
A.~D. Linde, {\it {Chaotic Inflation}},  {\em Phys.Lett.} {\bf B129} (1983)
  177--181.

\bibitem{Ade:2013uln}
{\bf Planck Collaboration} Collaboration, P.~Ade et~al., {\it {Planck 2013
  results. XXII. Constraints on inflation}},  {\em Astron.Astrophys.} {\bf 571}
  (2014) A22, [\href{http://arxiv.org/abs/1303.5082}{{\tt arXiv:1303.5082}}].

\bibitem{Planck:2015xua}
{\bf Planck Collaboration} Collaboration, {\it {Planck 2015 results. XIII.
  Cosmological parameters}},  \href{http://arxiv.org/abs/1502.0158}{{\tt
  arXiv:1502.0158}}.

\bibitem{Freese:1990rb}
K.~Freese, J.~A. Frieman, and A.~V. Olinto, {\it {Natural inflation with pseudo
  - Nambu-Goldstone bosons}},  {\em Phys.Rev.Lett.} {\bf 65} (1990) 3233--3236.

\bibitem{Silverstein:2008sg}
E.~Silverstein and A.~Westphal, {\it {Monodromy in the CMB: Gravity Waves and
  String Inflation}},  {\em Phys.Rev.} {\bf D78} (2008) 106003,
  [\href{http://arxiv.org/abs/0803.3085}{{\tt arXiv:0803.3085}}].

\bibitem{Berera:1995ie}
A.~Berera, {\it {Warm inflation}},  {\em Phys.Rev.Lett.} {\bf 75} (1995)
  3218--3221, [\href{http://arxiv.org/abs/astro-ph/9509049}{{\tt
  astro-ph/9509049}}].

\bibitem{Berera:1999ws}
A.~Berera, {\it {Warm inflation at arbitrary adiabaticity: A Model, an
  existence proof for inflationary dynamics in quantum field theory}},  {\em
  Nucl.Phys.} {\bf B585} (2000) 666--714,
  [\href{http://arxiv.org/abs/hep-ph/9904409}{{\tt hep-ph/9904409}}].

\bibitem{Lyth:1996im}
D.~H. Lyth, {\it {What would we learn by detecting a gravitational wave signal
  in the cosmic microwave background anisotropy?}},  {\em Phys.Rev.Lett.} {\bf
  78} (1997) 1861--1863, [\href{http://arxiv.org/abs/hep-ph/9606387}{{\tt
  hep-ph/9606387}}].

\bibitem{Efstathiou:2005tq}
G.~Efstathiou and K.~J. Mack, {\it {The Lyth bound revisited}},  {\em JCAP}
  {\bf 0505} (2005) 008, [\href{http://arxiv.org/abs/astro-ph/0503360}{{\tt
  astro-ph/0503360}}].

\bibitem{Easther:2006qu}
R.~Easther, W.~H. Kinney, and B.~A. Powell, {\it {The Lyth bound and the end of
  inflation}},  {\em JCAP} {\bf 0608} (2006) 004,
  [\href{http://arxiv.org/abs/astro-ph/0601276}{{\tt astro-ph/0601276}}].

\bibitem{Baumann:2011ws}
D.~Baumann and D.~Green, {\it {A Field Range Bound for General Single-Field
  Inflation}},  {\em JCAP} {\bf 1205} (2012) 017,
  [\href{http://arxiv.org/abs/1111.3040}{{\tt arXiv:1111.3040}}].

\bibitem{Antusch:2014cpa}
S.~Antusch and D.~Nolde, {\it {BICEP2 implications for single-field slow-roll
  inflation revisited}},  {\em JCAP} {\bf 1405} (2014) 035,
  [\href{http://arxiv.org/abs/1404.1821}{{\tt arXiv:1404.1821}}].

\bibitem{Garcia-Bellido:2014eva}
J.~Garcia-Bellido, D.~Roest, M.~Scalisi, and I.~Zavala, {\it {Can CMB data
  constrain the inflationary field range?}},  {\em JCAP} {\bf 1409} (2014) 006,
  [\href{http://arxiv.org/abs/1405.7399}{{\tt arXiv:1405.7399}}].

\bibitem{Gao:2014pca}
Q.~Gao, Y.~Gong, and T.~Li, {\it {The Modified Lyth Bound and Implications of
  BICEP2 Results}},  \href{http://arxiv.org/abs/1405.6451}{{\tt
  arXiv:1405.6451}}.

\bibitem{Bramante:2014rva}
J.~Bramante, S.~Downes, L.~Lehman, and A.~Martin, {\it {Clearing the Brush: The
  Last Stand of Solo Small Field Inflation}},  {\em Phys.Rev.} {\bf D90} (2014)
  023530, [\href{http://arxiv.org/abs/1405.7563}{{\tt arXiv:1405.7563}}].

\bibitem{Garcia-Bellido:2014wfa}
J.~Garcia-Bellido, D.~Roest, M.~Scalisi, and I.~Zavala, {\it {The Lyth Bound of
  Inflation with a Tilt}},  {\em Phys.Rev.} {\bf D90} (2014) 123539,
  [\href{http://arxiv.org/abs/1408.6839}{{\tt arXiv:1408.6839}}].

\bibitem{Ade:2014xna}
{\bf BICEP2 Collaboration} Collaboration, P.~Ade et~al., {\it {Detection of
  B-Mode Polarization at Degree Angular Scales by BICEP2}},  {\em
  Phys.Rev.Lett.} {\bf 112} (2014) 241101,
  [\href{http://arxiv.org/abs/1403.3985}{{\tt arXiv:1403.3985}}].

\bibitem{Dine:1995uk}
M.~Dine, L.~Randall, and S.~D. Thomas, {\it {Supersymmetry breaking in the
  early universe}},  {\em Phys.Rev.Lett.} {\bf 75} (1995) 398--401,
  [\href{http://arxiv.org/abs/hep-ph/9503303}{{\tt hep-ph/9503303}}].

\bibitem{Gherghetta:1995dv}
T.~Gherghetta, C.~F. Kolda, and S.~P. Martin, {\it {Flat directions in the
  scalar potential of the supersymmetric standard model}},  {\em Nucl.Phys.}
  {\bf B468} (1996) 37--58, [\href{http://arxiv.org/abs/hep-ph/9510370}{{\tt
  hep-ph/9510370}}].

\bibitem{Allahverdi:2006iq}
R.~Allahverdi, K.~Enqvist, J.~Garcia-Bellido, and A.~Mazumdar, {\it {Gauge
  invariant MSSM inflaton}},  {\em Phys.Rev.Lett.} {\bf 97} (2006) 191304,
  [\href{http://arxiv.org/abs/hep-ph/0605035}{{\tt hep-ph/0605035}}].

\bibitem{Allahverdi:2006we}
R.~Allahverdi, K.~Enqvist, J.~Garcia-Bellido, A.~Jokinen, and A.~Mazumdar, {\it
  {MSSM flat direction inflation: Slow roll, stability, fine tunning and
  reheating}},  {\em JCAP} {\bf 0706} (2007) 019,
  [\href{http://arxiv.org/abs/hep-ph/0610134}{{\tt hep-ph/0610134}}].

\bibitem{Enqvist:2007tf}
K.~Enqvist, L.~Mether, and S.~Nurmi, {\it {Supergravity origin of the MSSM
  inflation}},  {\em JCAP} {\bf 0711} (2007) 014,
  [\href{http://arxiv.org/abs/0706.2355}{{\tt arXiv:0706.2355}}].

\bibitem{Allahverdi:2011su}
R.~Allahverdi, S.~Downes, and B.~Dutta, {\it {Constructing Flat Inflationary
  Potentials in Supersymmetry}},  {\em Phys.Rev.} {\bf D84} (2011) 101301,
  [\href{http://arxiv.org/abs/1106.5004}{{\tt arXiv:1106.5004}}].

\bibitem{Yamada:2012tj}
Y.~Yamada, {\it {Instant uplifted inflation: A solution for a tension between
  inflation and SUSY breaking scale}},  {\em JHEP} {\bf 1307} (2013) 039,
  [\href{http://arxiv.org/abs/1211.1455}{{\tt arXiv:1211.1455}}].

\bibitem{Mollerach:1989hu}
S.~Mollerach, {\it {Isocurvature Baryon Perturbations and Inflation}},  {\em
  Phys.Rev.} {\bf D42} (1990) 313--325.

\bibitem{Linde:1996gt}
A.~D. Linde and V.~F. Mukhanov, {\it {Nongaussian isocurvature perturbations
  from inflation}},  {\em Phys.Rev.} {\bf D56} (1997) 535--539,
  [\href{http://arxiv.org/abs/astro-ph/9610219}{{\tt astro-ph/9610219}}].

\bibitem{Lyth:2001nq}
D.~H. Lyth and D.~Wands, {\it {Generating the curvature perturbation without an
  inflaton}},  {\em Phys.Lett.} {\bf B524} (2002) 5--14,
  [\href{http://arxiv.org/abs/hep-ph/0110002}{{\tt hep-ph/0110002}}].

\bibitem{Kodama:1985bj}
H.~Kodama and M.~Sasaki, {\it {Cosmological Perturbation Theory}},  {\em
  Prog.Theor.Phys.Suppl.} {\bf 78} (1984) 1--166.

\bibitem{Gordon:2000hv}
C.~Gordon, D.~Wands, B.~A. Bassett, and R.~Maartens, {\it {Adiabatic and
  entropy perturbations from inflation}},  {\em Phys.Rev.} {\bf D63} (2001)
  023506, [\href{http://arxiv.org/abs/astro-ph/0009131}{{\tt
  astro-ph/0009131}}].

\bibitem{Enqvist:2011pt}
K.~Enqvist, D.~G. Figueroa, and G.~Rigopoulos, {\it {Fluctuations along
  supersymmetric flat directions during Inflation}},  {\em JCAP} {\bf 1201}
  (2012) 053, [\href{http://arxiv.org/abs/1109.3024}{{\tt arXiv:1109.3024}}].

\bibitem{Byrnes:2008wi}
C.~T. Byrnes, K.-Y. Choi, and L.~M. Hall, {\it {Conditions for large
  non-Gaussianity in two-field slow-roll inflation}},  {\em JCAP} {\bf 0810}
  (2008) 008, [\href{http://arxiv.org/abs/0807.1101}{{\tt arXiv:0807.1101}}].

\bibitem{Gangui:1993tt}
A.~Gangui, F.~Lucchin, S.~Matarrese, and S.~Mollerach, {\it {The Three point
  correlation function of the cosmic microwave background in inflationary
  models}},  {\em Astrophys.J.} {\bf 430} (1994) 447--457,
  [\href{http://arxiv.org/abs/astro-ph/9312033}{{\tt astro-ph/9312033}}].

\bibitem{Maldacena:2002vr}
J.~M. Maldacena, {\it {Non-Gaussian features of primordial fluctuations in
  single field inflationary models}},  {\em JHEP} {\bf 0305} (2003) 013,
  [\href{http://arxiv.org/abs/astro-ph/0210603}{{\tt astro-ph/0210603}}].

\bibitem{Vernizzi:2006ve}
F.~Vernizzi and D.~Wands, {\it {Non-gaussianities in two-field inflation}},
  {\em JCAP} {\bf 0605} (2006) 019,
  [\href{http://arxiv.org/abs/astro-ph/0603799}{{\tt astro-ph/0603799}}].

\bibitem{Lewis:2011au}
A.~Lewis, {\it {The real shape of non-Gaussianities}},  {\em JCAP} {\bf 1110}
  (2011) 026, [\href{http://arxiv.org/abs/1107.5431}{{\tt arXiv:1107.5431}}].

\bibitem{Peterson:2010mv}
C.~M. Peterson and M.~Tegmark, {\it {Non-Gaussianity in Two-Field Inflation}},
  {\em Phys.Rev.} {\bf D84} (2011) 023520,
  [\href{http://arxiv.org/abs/1011.6675}{{\tt arXiv:1011.6675}}].

\bibitem{Tanaka:2010km}
T.~Tanaka, T.~Suyama, and S.~Yokoyama, {\it {Use of delta N formalism -
  Difficulties in generating large local-type non-Gaussianity during
  inflation}},  {\em Class.Quant.Grav.} {\bf 27} (2010) 124003,
  [\href{http://arxiv.org/abs/1003.5057}{{\tt arXiv:1003.5057}}].

\bibitem{Gong:2011cd}
J.-O. Gong and H.~M. Lee, {\it {Large non-Gaussianity in non-minimal
  inflation}},  {\em JCAP} {\bf 1111} (2011) 040,
  [\href{http://arxiv.org/abs/1105.0073}{{\tt arXiv:1105.0073}}].

\bibitem{Elliston:2011dr}
J.~Elliston, D.~J. Mulryne, D.~Seery, and R.~Tavakol, {\it {Evolution of fNL to
  the adiabatic limit}},  {\em JCAP} {\bf 1111} (2011) 005,
  [\href{http://arxiv.org/abs/1106.2153}{{\tt arXiv:1106.2153}}].

\bibitem{Elliston:2011et}
J.~Elliston, D.~Mulryne, D.~Seery, and R.~Tavakol, {\it {Evolution of
  non-Gaussianity in multi-scalar field models}},  {\em Int.J.Mod.Phys.} {\bf
  A26} (2011) 3821--3832, [\href{http://arxiv.org/abs/1107.2270}{{\tt
  arXiv:1107.2270}}].

\bibitem{Choi:2011me}
K.-Y. Choi and B.~Kyae, {\it {Natural Hybrid Inflation Model with Large
  Non-Gaussianity}},  {\em Phys.Lett.} {\bf B706} (2012) 243--250,
  [\href{http://arxiv.org/abs/1109.4245}{{\tt arXiv:1109.4245}}].

\bibitem{Mazumdar:2012jj}
A.~Mazumdar and L.-F. Wang, {\it {Separable and non-separable multi-field
  inflation and large non-Gaussianity}},  {\em JCAP} {\bf 1209} (2012) 005,
  [\href{http://arxiv.org/abs/1203.3558}{{\tt arXiv:1203.3558}}].

\bibitem{Lyth:2004gb}
D.~H. Lyth, K.~A. Malik, and M.~Sasaki, {\it {A General proof of the
  conservation of the curvature perturbation}},  {\em JCAP} {\bf 0505} (2005)
  004, [\href{http://arxiv.org/abs/astro-ph/0411220}{{\tt astro-ph/0411220}}].

\bibitem{Sasaki:1995aw}
M.~Sasaki and E.~D. Stewart, {\it {A General analytic formula for the spectral
  index of the density perturbations produced during inflation}},  {\em
  Prog.Theor.Phys.} {\bf 95} (1996) 71--78,
  [\href{http://arxiv.org/abs/astro-ph/9507001}{{\tt astro-ph/9507001}}].

\bibitem{Sasaki:1998ug}
M.~Sasaki and T.~Tanaka, {\it {Superhorizon scale dynamics of multiscalar
  inflation}},  {\em Prog.Theor.Phys.} {\bf 99} (1998) 763--782,
  [\href{http://arxiv.org/abs/gr-qc/9801017}{{\tt gr-qc/9801017}}].

\bibitem{Wands:2000dp}
D.~Wands, K.~A. Malik, D.~H. Lyth, and A.~R. Liddle, {\it {A New approach to
  the evolution of cosmological perturbations on large scales}},  {\em
  Phys.Rev.} {\bf D62} (2000) 043527,
  [\href{http://arxiv.org/abs/astro-ph/0003278}{{\tt astro-ph/0003278}}].

\bibitem{Lyth:2005fi}
D.~H. Lyth and Y.~Rodriguez, {\it {The Inflationary prediction for primordial
  non-Gaussianity}},  {\em Phys.Rev.Lett.} {\bf 95} (2005) 121302,
  [\href{http://arxiv.org/abs/astro-ph/0504045}{{\tt astro-ph/0504045}}].

\bibitem{Choi:2007su}
K.-Y. Choi, L.~M. Hall, and C.~van~de Bruck, {\it {Spectral Running and
  Non-Gaussianity from Slow-Roll Inflation in Generalised Two-Field Models}},
  {\em JCAP} {\bf 0702} (2007) 029,
  [\href{http://arxiv.org/abs/astro-ph/0701247}{{\tt astro-ph/0701247}}].

\bibitem{Lyth:2005qj}
D.~H. Lyth and I.~Zaballa, {\it {A Bound concerning primordial
  non-Gaussianity}},  {\em JCAP} {\bf 0510} (2005) 005,
  [\href{http://arxiv.org/abs/astro-ph/0507608}{{\tt astro-ph/0507608}}].

\bibitem{Enqvist:2010vd}
K.~Enqvist, A.~Mazumdar, and P.~Stephens, {\it {Inflection point inflation
  within supersymmetry}},  {\em JCAP} {\bf 1006} (2010) 020,
  [\href{http://arxiv.org/abs/1004.3724}{{\tt arXiv:1004.3724}}].

\bibitem{Byrnes:2008zy}
C.~T. Byrnes, K.-Y. Choi, and L.~M. Hall, {\it {Large non-Gaussianity from
  two-component hybrid inflation}},  {\em JCAP} {\bf 0902} (2009) 017,
  [\href{http://arxiv.org/abs/0812.0807}{{\tt arXiv:0812.0807}}].

\bibitem{Krause:2007jk}
A.~Krause and E.~Pajer, {\it {Chasing brane inflation in string-theory}},  {\em
  JCAP} {\bf 0807} (2008) 023, [\href{http://arxiv.org/abs/0705.4682}{{\tt
  arXiv:0705.4682}}].

\bibitem{Baumann:2007np}
D.~Baumann, A.~Dymarsky, I.~R. Klebanov, L.~McAllister, and P.~J. Steinhardt,
  {\it {A Delicate universe}},  {\em Phys.Rev.Lett.} {\bf 99} (2007) 141601,
  [\href{http://arxiv.org/abs/0705.3837}{{\tt arXiv:0705.3837}}].

\bibitem{Baumann:2007ah}
D.~Baumann, A.~Dymarsky, I.~R. Klebanov, and L.~McAllister, {\it {Towards an
  Explicit Model of D-brane Inflation}},  {\em JCAP} {\bf 0801} (2008) 024,
  [\href{http://arxiv.org/abs/0706.0360}{{\tt arXiv:0706.0360}}].

\bibitem{Baumann:2008kq}
D.~Baumann, A.~Dymarsky, S.~Kachru, I.~R. Klebanov, and L.~McAllister, {\it
  {Holographic Systematics of D-brane Inflation}},  {\em JHEP} {\bf 0903}
  (2009) 093, [\href{http://arxiv.org/abs/0808.2811}{{\tt arXiv:0808.2811}}].

\bibitem{Burgess:2008ir}
C.~Burgess, J.~M. Cline, and M.~Postma, {\it {Axionic D3-D7 Inflation}},  {\em
  JHEP} {\bf 0903} (2009) 058, [\href{http://arxiv.org/abs/0811.1503}{{\tt
  arXiv:0811.1503}}].

\bibitem{Allahverdi:2008bt}
R.~Allahverdi, B.~Dutta, and A.~Mazumdar, {\it {Attraction towards an
  inflection point inflation}},  {\em Phys.Rev.} {\bf D78} (2008) 063507,
  [\href{http://arxiv.org/abs/0806.4557}{{\tt arXiv:0806.4557}}].

\bibitem{Chen:2009nk}
H.-Y. Chen, L.-Y. Hung, and G.~Shiu, {\it {Inflation on an Open Racetrack}},
  {\em JHEP} {\bf 0903} (2009) 083, [\href{http://arxiv.org/abs/0901.0267}{{\tt
  arXiv:0901.0267}}].

\bibitem{Badziak:2009eh}
M.~Badziak and M.~Olechowski, {\it {Inflation with racetrack superpotential and
  matter field}},  {\em JCAP} {\bf 1002} (2010) 026,
  [\href{http://arxiv.org/abs/0911.1213}{{\tt arXiv:0911.1213}}].

\bibitem{Agarwal:2011wm}
N.~Agarwal, R.~Bean, L.~McAllister, and G.~Xu, {\it {Universality in D-brane
  Inflation}},  {\em JCAP} {\bf 1109} (2011) 002,
  [\href{http://arxiv.org/abs/1103.2775}{{\tt arXiv:1103.2775}}].

\bibitem{Downes:2011gi}
S.~Downes, B.~Dutta, and K.~Sinha, {\it {Catastrophic Inflation}},  {\em
  Phys.Rev.} {\bf D84} (2011) 063524,
  [\href{http://arxiv.org/abs/1106.2266}{{\tt arXiv:1106.2266}}].

\bibitem{Elliston:2012wm}
J.~Elliston, L.~Alabidi, I.~Huston, D.~Mulryne, and R.~Tavakol, {\it {Large
  trispectrum in two-field slow-roll inflation}},  {\em JCAP} {\bf 1209} (2012)
  001, [\href{http://arxiv.org/abs/1203.6844}{{\tt arXiv:1203.6844}}].

\bibitem{Downes:2012xb}
S.~Downes, B.~Dutta, and K.~Sinha, {\it {Attractors, Universality and
  Inflation}},  {\em Phys.Rev.} {\bf D86} (2012) 103509,
  [\href{http://arxiv.org/abs/1203.6892}{{\tt arXiv:1203.6892}}].

\bibitem{McAllister:2012am}
L.~McAllister, S.~Renaux-Petel, and G.~Xu, {\it {A Statistical Approach to
  Multifield Inflation: Many-field Perturbations Beyond Slow Roll}},  {\em
  JCAP} {\bf 1210} (2012) 046, [\href{http://arxiv.org/abs/1207.0317}{{\tt
  arXiv:1207.0317}}].

\bibitem{Erdmenger:2012gx}
J.~Erdmenger, S.~Halter, C.~Nunez, and G.~Tasinato, {\it {Slow-walking
  inflation}},  {\em JCAP} {\bf 1301} (2013) 006,
  [\href{http://arxiv.org/abs/1210.4179}{{\tt arXiv:1210.4179}}].

\bibitem{Cerezo:2012ub}
R.~Cerezo and J.~G. Rosa, {\it {Warm Inflection}},  {\em JHEP} {\bf 1301}
  (2013) 024, [\href{http://arxiv.org/abs/1210.7975}{{\tt arXiv:1210.7975}}].

\bibitem{Downes:2012gu}
S.~Downes and B.~Dutta, {\it {Inflection Points and the Power Spectrum}},  {\em
  Phys.Rev.} {\bf D87} (2013), no.~8 083518,
  [\href{http://arxiv.org/abs/1211.1707}{{\tt arXiv:1211.1707}}].

\bibitem{Choudhury:2013jya}
S.~Choudhury, A.~Mazumdar, and S.~Pal, {\it {Low and High scale MSSM inflation,
  gravitational waves and constraints from Planck}},  {\em JCAP} {\bf 1307}
  (2013) 041, [\href{http://arxiv.org/abs/1305.6398}{{\tt arXiv:1305.6398}}].

\bibitem{Pedro:2013pba}
F.~G. Pedro and A.~Westphal, {\it {Low-$\ell$ CMB power loss in string
  inflation}},  {\em JHEP} {\bf 1404} (2014) 034,
  [\href{http://arxiv.org/abs/1309.3413}{{\tt arXiv:1309.3413}}].

\end{thebibliography}\endgroup

\end{document}